\documentclass[3p,times,twocolumn]{elsarticle}
\usepackage{hyperref,subfig,float,siunitx,upgreek,makecell,textcomp}

\journal{Nuclear and Particle Physics Proceedings}

\begin{document}

\begin{frontmatter}
\title{Lifetime and Performance of the very latest Microchannel-Plate Photomultipliers}
\author[erlangen]{{D. Miehling}\corref{mycorrespondingauthor}}
\cortext[mycorrespondingauthor]{Corresponding author}
\ead{daniel.miehling@fau.de}

\author[erlangen]{M. B\"ohm}
\author[erlangen]{K. Gumbert}
\author[erlangen]{S. Krauss}
\author[erlangen]{A. Lehmann}

\author[gsi]{A.~Belias}
\author[gsi]{R.~Dzhygadlo}
\author[gsi]{A.~Gerhardt}
\author[gsi]{D.~Lehmann}
\author[gsi,ff]{K.~Peters}
\author[gsi]{G.~Schepers}
\author[gsi]{C.~Schwarz}
\author[gsi]{J.~Schwiening}
\author[gsi]{M.~Traxler}
\author[gsi,ff]{Y.~Wolf}  

\author[fa]{L.~Schmitt} 

\author[giessen]{M. D\"uren}
\author[giessen]{A. Hayrapetyan}
\author[giessen]{I. K\"oseoglu}
\author[giessen]{M. Schmidt}
\author[giessen]{T. Wasem}

\author[mainz]{C. Sfienti}

\author[him]{A. Ali}

\address[erlangen]{Friedrich
	Alexander-University of Erlangen-Nuremberg, Erlangen, Germany}
\address[gsi]{GSI Helmholtzzentrum f\"ur Schwerionenforschung GmbH , Darmstadt, Germany}
\address[ff]{Goethe-University, Frankfurt, Germany} 
\address[giessen]{II. Physikalisches Institut, Justus Liebig-University of Giessen, Giessen, Germany}
\address[fa]{FAIR, Facility for Antiproton and Ion Research in Europe, Darmstadt, Germany}
\address[mainz]{Institut f\"ur Kernphysik, Johannes Gutenberg-University of Mainz, Mainz, Germany}
\address[him]{Helmholtz-Institut Mainz, Germany}

\begin{abstract}
The PANDA experiment at the FAIR facility at GSI will study hadron physics using a high intensity antiproton beam of up to 15 GeV/c momentum to perform high precision spectroscopy. Two DIRC detectors with their image planes residing in an $\sim$1$\,$T magnetic field will be used in the experiment. The only suitable photon detectors for both DIRCs were identified to be Microchannel-Plate Photomultipliers (MCP-PMTs). Since the aging problems of MCP-PMTs were solved recently by coating the MCPs with the so-called ALD-technique (atomic layer depostion) we are investigating devices which are significantly improved with respect to other parameters, as, e.g., the collection efficiency (CE) and the quantum efficiency (QE). The latest generation of MCP-PMTs can reach a detective quantum efficiency DQE$\,=\,$QE\,$\cdot$\,CE of ~30\%. This paper will present the performance of the most advanced 53x53 mm$^2$ ALD-coated MCP-PMTs from Photonis (8x8 and 3x100 anodes) and Photek (8x8 anodes), also inside the magnetic field. With a picosecond laser and a multi-hit capable DAQ system which allows read out up to 300 pixels simultaneously, parameters like darkcount rate, afterpulse probability and time resolution can be investigated as a function of incident photon position.
\end{abstract}


\end{frontmatter}


\section{Introduction}

In the PANDA experiment \cite{pandatpr,pandappr,BADtdr,EDDTDR} two DIRC (Detection of Internally Reflected Chrenkov light) detectors will be used for particle identification. For both of them the focal plane will reside in a magnetic field of $\sim$1$\,$T. This and some other constraints like darkcount rate and radiation hardness lead to microchannel-plate photomultipliers (MCP-PMTs) as the only suitable photodetector option \cite{coyle,Dzhygadlo,DUREN2017}. They have a low darkcount rate of ${<}\,1\,\textrm{kHz}/\textrm{cm}^2$ and a high time resolution of less than $50\,\textrm{ps}$ $\sigma_{TTS}$ and $500\,\textrm{ps}$ $\sigma_{RMS}$. These can detect single photons, withstand the high photon rates of the Barrel DIRC (${\sim} 200\,\textrm{kHz}/\textrm{cm}^2$) and the Endcap Disc DIRC (EDD, ${\sim} 1\,\textrm{MHz}/\textrm{cm}^2$) and are compact enough to fit the space limitations of both of the DIRC detectors. Most importantly, they can withstand the existing magnetic field.

A long-standing problem with MCP-PMTs was their low lifetime. After a few hundred $\textrm{mC}/\textrm{cm}^2$ of integrated anode charge (IAC) the quantum efficiency (QE) had dropped by a factor of 2 or more \cite{Kishimoto:2006mg,Britting:2011zz,Uhlig:2012rk}. This aging is caused by heavy feedback ions from the residual gas \cite{EDIT_2010}. 
With the above mentioned photon rates, a runtime of 10 years, 50\% duty cycle, and a $10^6$ gain for the MCP-PMTs an IAC of ${\sim} 5\,\textrm{C}/\textrm{cm}^2$ for the Barrel DIRC and even more for the EDD is expected. With the development of the atomic layer deposition (ALD) technique \cite{BEAULIEU200981} these values are now reachable. In this process the MCPs are coated with a thin layer of alumina or magnesia to effectively prevent outgassing of the MCPs which reduces the feedback ion flux significantly and leads to a much higher lifetime. The first ALD-coated MCPs were produced by Arradiance Inc., now the technique is used by Photonis, the LAPPD collaboration \cite{WETSTEIN2011148,SIEGMUND2012168}, Hamamatsu and Photek \cite{CONNEELY2013388}.

Another major recent improvement was an increase of the collection efficiency (CE) from $\sim$65\% to more than 90\% which led alongside with great improvements in the quantum efficiency to a overall detection efficiency of $\sim$30\% (DQE = QE $\cdot$ CE) instead of $ \sim$12\% for former tubes. This high CE is available in the latest tubes of Photonis and Photek as will be shown in this paper.

\section{Measurements of various parameters and their results}
In the following section an overview of different performance parameters and the results reached by the latest ALD-coated 53x53$\,$mm$^2$ MCP-PMTs of Photek and Photonis will be given. Both manufacturers use two MCPs in chevron configuration and gaps of a few 100\,\textmu m in their sensors to reach a compact design. Typical resistances range from 10 to 100\,M$\Omega$.
\subsection{Quantum efficiency and lifetime}
\begin{figure}[h]
	\centering
		
			\includegraphics[width=0.4\textwidth]{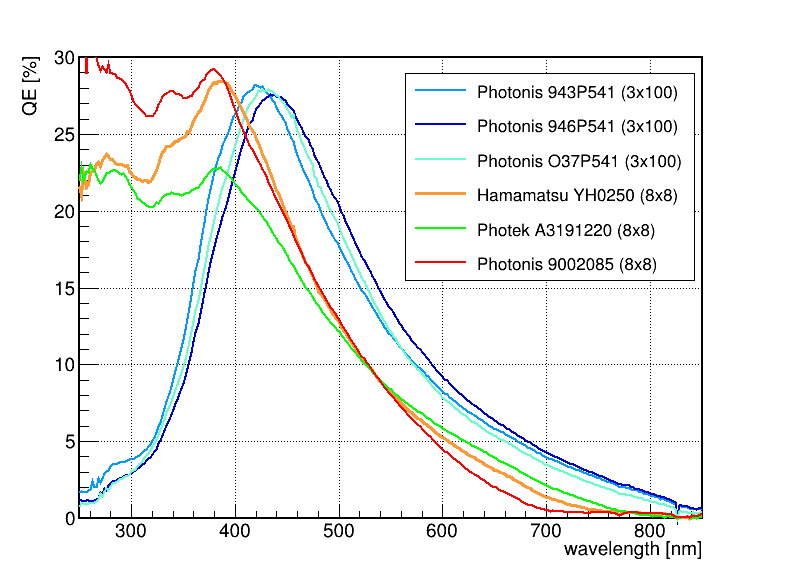}
			
			\caption{comparison of wavelength dependend QE}
			\label{fig:qewave}
\end{figure}

The quantum efficiency is measured in two ways in Erlangen. First, as a function of the wavelength at one position and second position dependent across the whole sensor surface at a certain wavelength (laer wavelength 372$\,$nm). Figure \ref{fig:qewave} shows the spectral QE of different MCP-PMTs. All manufacturers can reach QEs of well above 20\% (Photek in green) or even almost 30\% (Hamamatsu in orange or Photonis in red). It is also possible to tune the sensitivity of the photo cathode around a wavelength window (400-420$\,$nm in this case, Photonis in blue shades).

\begin{figure}[h]
	\centering	
			
			\subfloat[QE-Scans of Photonis 9002192 at $0\,\textrm{C}/\textrm{cm}^2$ (left) and $4.8\,\textrm{C}/\textrm{cm}^2$ (right)]{	
					\includegraphics[trim = 0mm 0mm 0mm 20mm, clip =true, width=0.20\textwidth ]{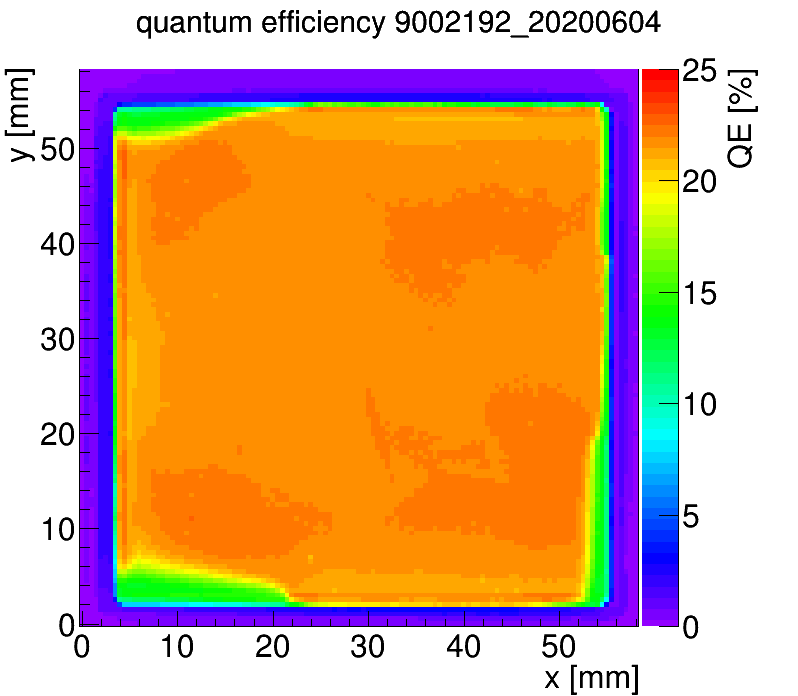}
					\includegraphics[trim = 0mm 0mm 0mm 20mm, clip =true,width=0.20\textwidth]{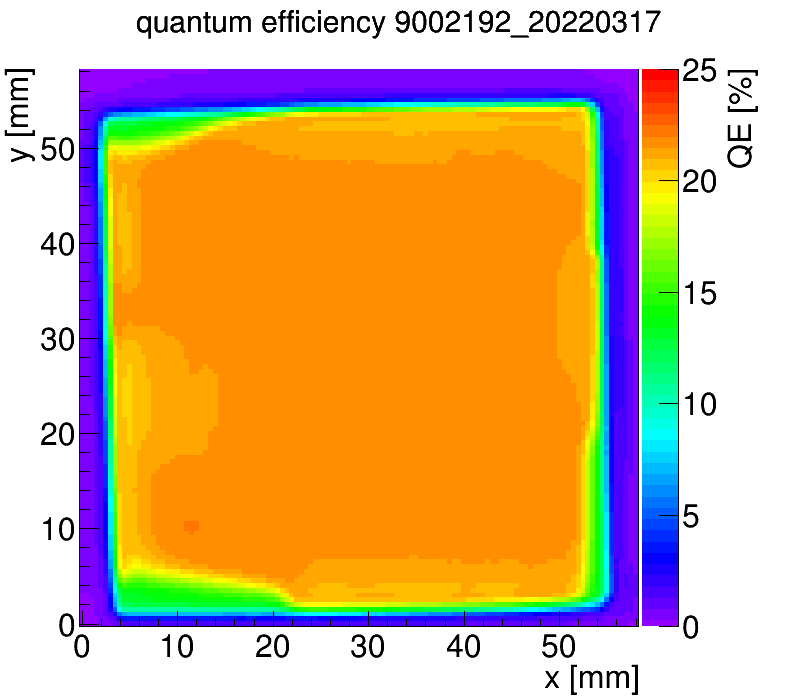}

		\label{fig:qescancompared_photonisa}}

			\subfloat[QE-Scans of Photonis 9002193 at $0\,\textrm{C}/\textrm{cm}^2$ (left) and $5\,\textrm{C}/\textrm{cm}^2$ (right)]{	
			\includegraphics[trim = 0mm 0mm 0mm 20mm, clip =true,width=0.20\textwidth]{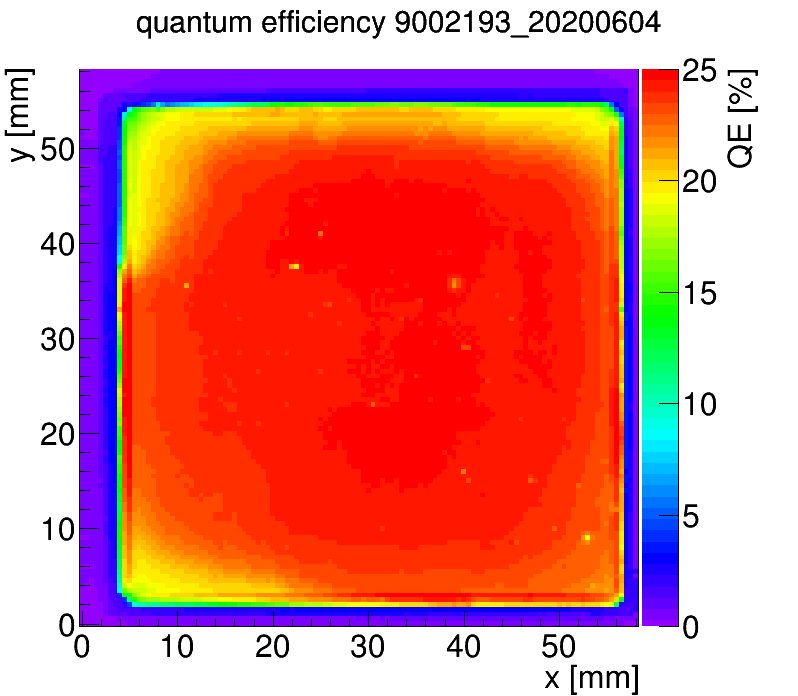}
			\includegraphics[trim = 0mm 0mm 0mm 20mm, clip =true,width=0.20\textwidth]{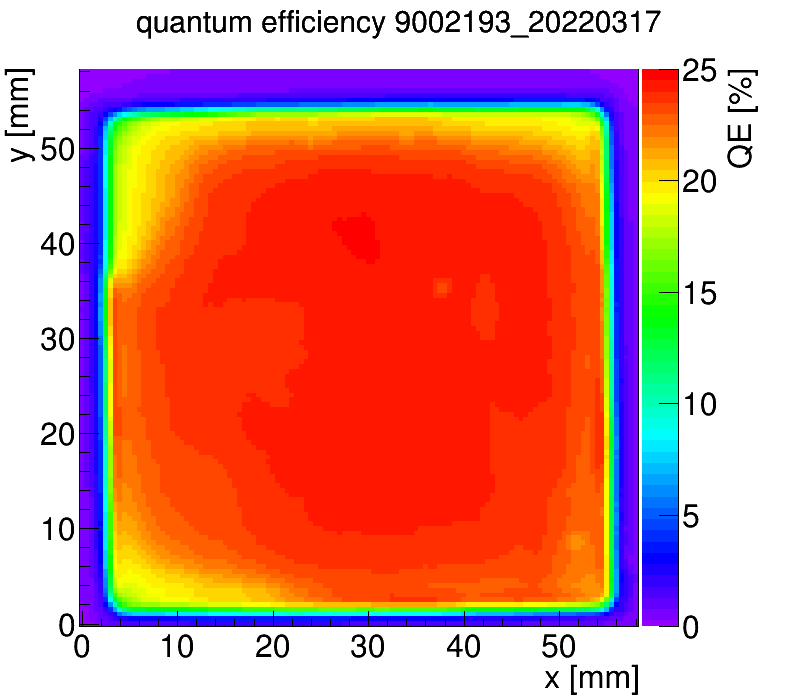}

		\label{fig:qescancompared_photonisb}}
		
			\subfloat[QE-Scans of Photonis 9001393 at $1.9\,\textrm{C}/\textrm{cm}^2$ (left) and $32\,\textrm{C}/\textrm{cm}^2$ (right)]{	
			\includegraphics[trim = 0mm 0mm 0mm 20mm, clip =true,width=0.20\textwidth]{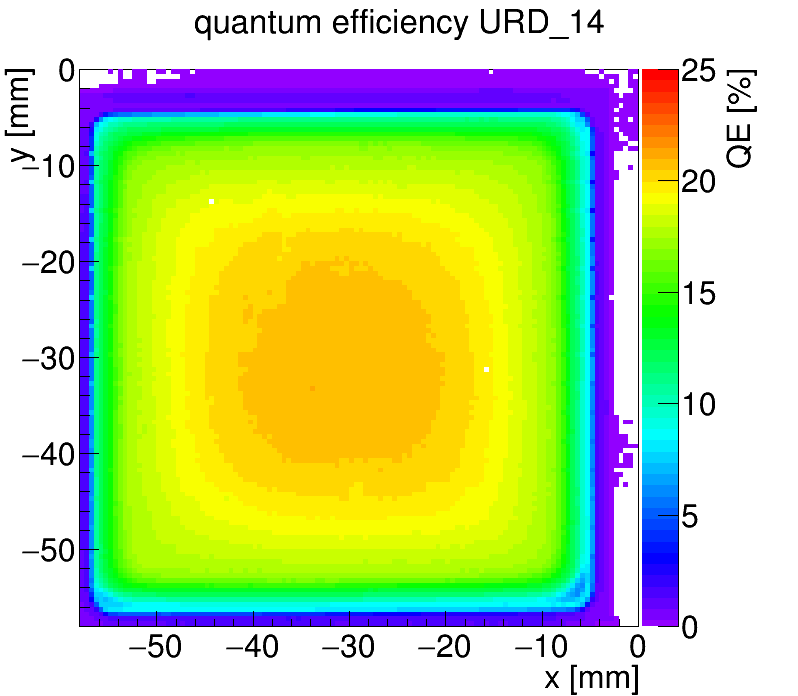}
			\includegraphics[trim = 0mm 0mm 0mm 20mm, clip =true,width=0.20\textwidth]{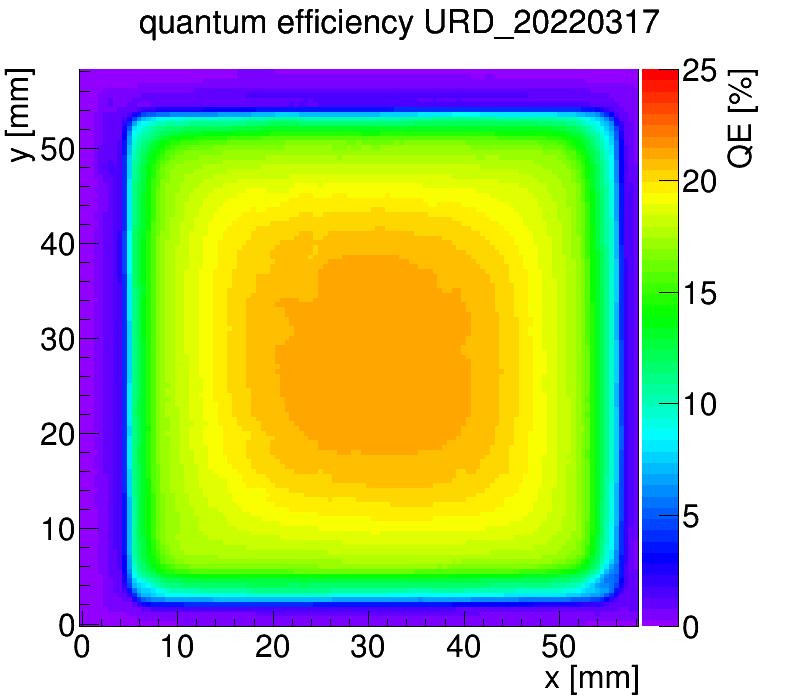}

		\label{fig:qescancompared_photonisc}}

		\caption{QE-Scans of Photonis MCP-PMTs at different IACs}	
		\label{fig:qescancompared_photonisd}

\end{figure}

\begin{figure}[h!]
	\centering	
			
			\subfloat[QE-Scans of Photek A3191220, at $0\,\textrm{C}/\textrm{cm}^2$ (left) and $1.4\,\textrm{C}/\textrm{cm}^2$ (right)]{	
			\includegraphics[trim = 0mm 0mm 0mm 17mm, clip =true, width=0.20\textwidth ]{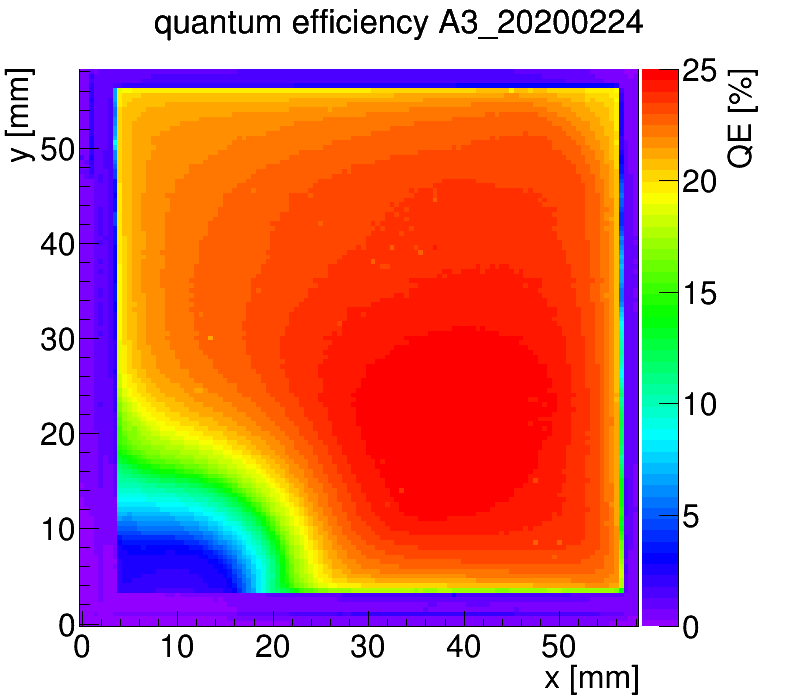}
			\includegraphics[trim = 0mm 0mm 0mm 17mm, clip =true,width=0.20\textwidth]{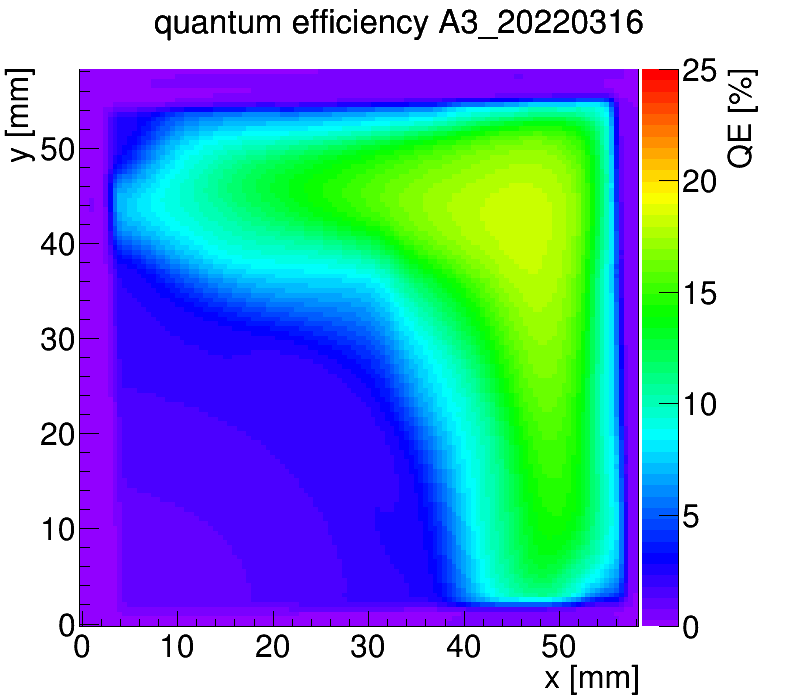}

		\label{fig:qescancompared_photeka}}

			\subfloat[QE-Scans of Photek A1200107, at $0\,\textrm{C}/\textrm{cm}^2$ (left) and $1.4\,\textrm{C}/\textrm{cm}^2$ (right)]{	
			\includegraphics[trim = 0mm 0mm 0mm 17mm, clip =true, width=0.20\textwidth ]{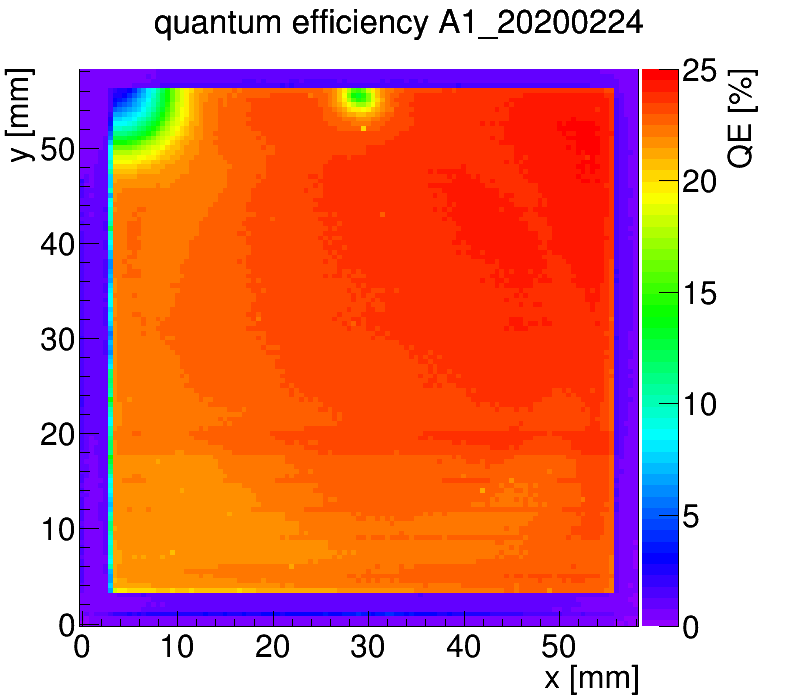}
			\includegraphics[trim = 0mm 0mm 0mm 17mm, clip =true,width=0.20\textwidth]{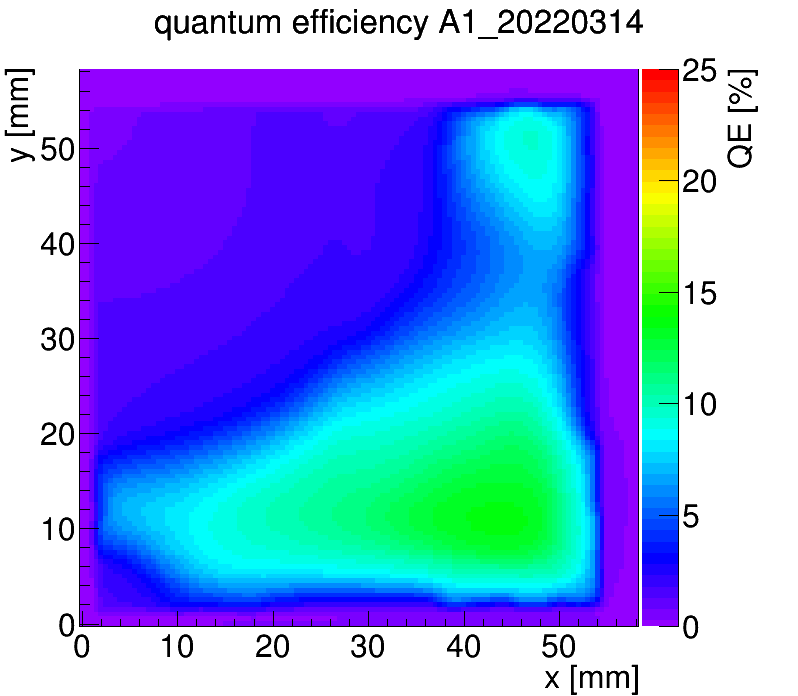}

		\label{fig:qescancompared_photekb}}

					\subfloat[QE-Scans of Photek A1200116 at $0\,\textrm{C}/\textrm{cm}^2$ (left) and still $0\,\textrm{C}/\textrm{cm}^2$ (right)]{
					
					\includegraphics[trim = 0mm 0mm 0mm 17mm, clip =true,width=0.20\textwidth]{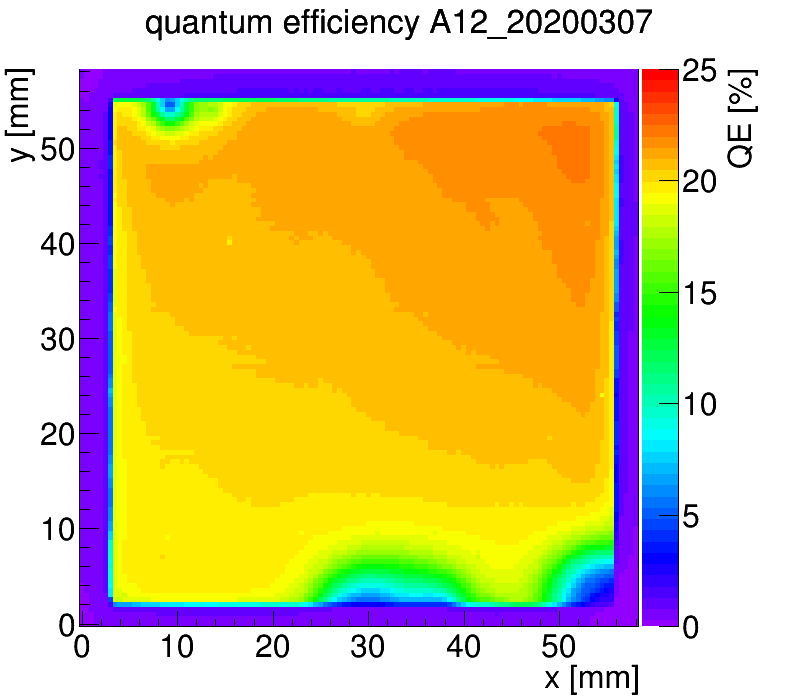}	
					
					\includegraphics[trim = 0mm 0mm 0mm 17mm, clip =true,width=0.20\textwidth]{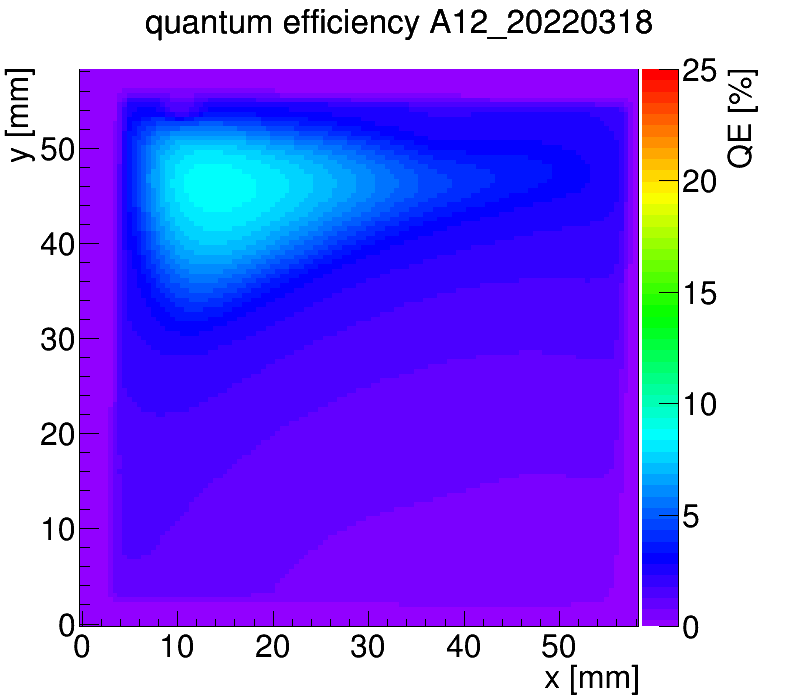}

		\label{fig:qescancompared_photekc}}
		
			\subfloat[QE-Scans of Photek A2200606 at $0\,\textrm{C}/\textrm{cm}^2$ (left) and $1.6\,\textrm{C}/\textrm{cm}^2$ (right)]{
			
					\includegraphics[trim = 0mm 0mm 0mm 17mm, clip =true,width=0.20\textwidth,]{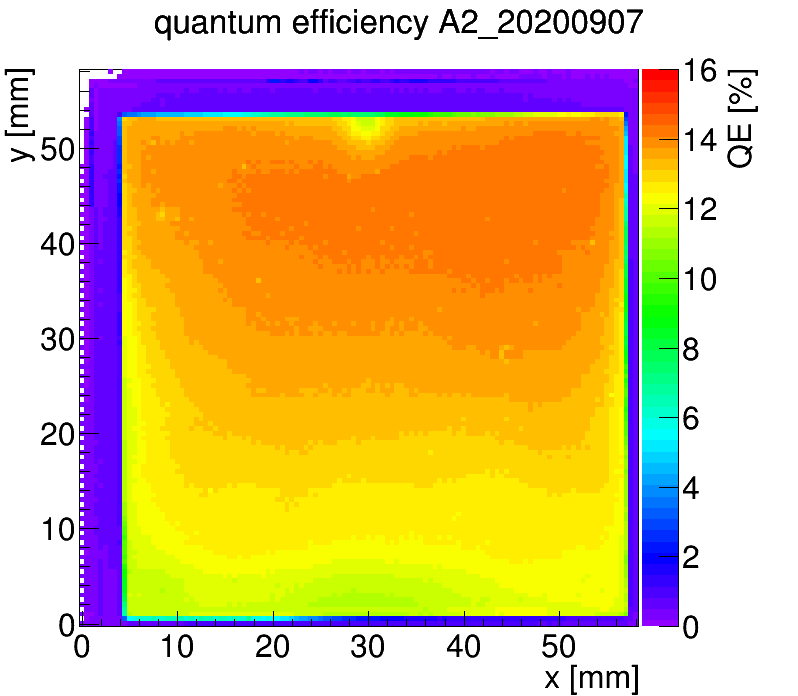}
					\includegraphics[trim = 0mm 0mm 0mm 17mm, clip =true,width=0.20\textwidth]{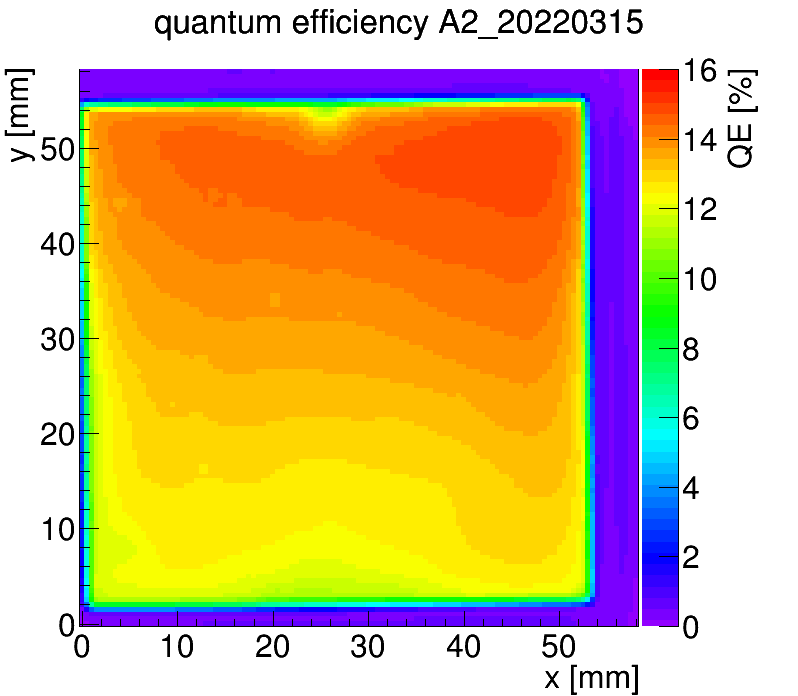}

		\label{fig:qescancompared_photekd}}

		\caption{QE-Scans of Photonis MCP-PMTs at different IACs}	
		\label{fig:qescancompared_photeke}

\end{figure}

\begin{figure}[h]
	\centering
	\includegraphics[width=0.5\textwidth]{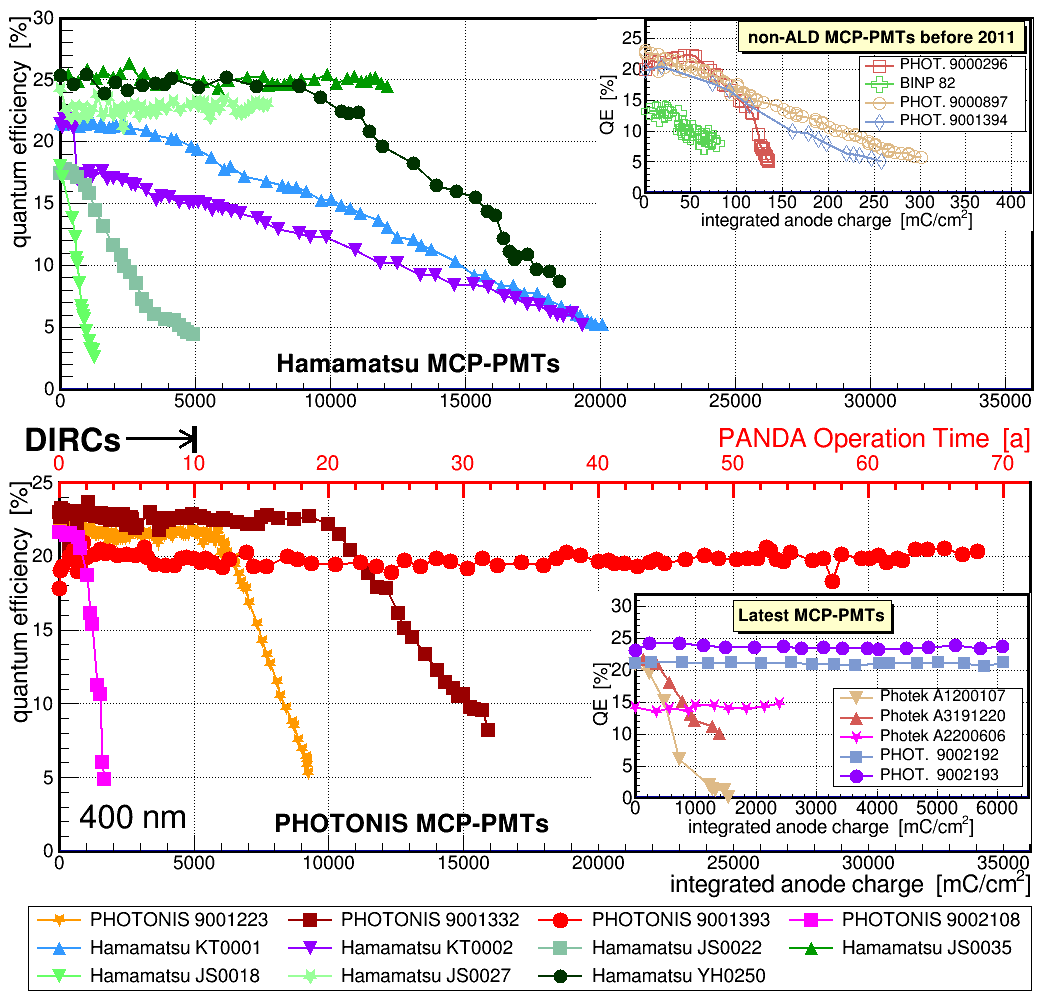}	
	\caption{results of various ALD-coated MCP-PMTs from Hamamatsu,  PHOTONIS, and Photek compared to some older non-ALD PMTs}
	\label{fig:lifetime}
\end{figure}

To measure the lifetime of a tube it is mounted in a setup where it is under constant illumination of $\sim$1$\,$MHz single photon rate (460$\,$nm) and the IAC is monitored. Every few weeks the tubes' spectral QE and every few months their position dependent QE is measured \cite{HEROLD2011S151}. In figure \ref{fig:qescancompared_photonisd} the spatial quantum efficiency for different IACs of three different Photonis sensors is shown. One can see no QE degradation yet.
In figure \ref{fig:qescancompared_photeke} the same is shown for some Photek tubes. The very left tube A3191220 was covered on the right side during illumination, the second tube A1200107 was covered on the left side. The third tube A1200116 was not under illumination at all, the shown data are just 2 years apart (March 2020 and March 2022). The last tube A2200606 is covered on the right side. One can see that the very first Photek tubes had a decreasing QE from the beginning even without illumination, probably caused by vacuum microleaks but the problem seems to be solved with A2200606 (although it has a low maximum QE of $\sim$15\%).
Figure \ref{fig:lifetime} shows the overall lifetime data for various MCP-PMTs. The by far best performing sensor is the Photonis 9001393 with 2 ALD-layers and an IAC of $>34\,\textrm{C}/\textrm{cm}^2$ with no QE loss yet.

\subsection{Collection efficiency and time resolution}
The collection efficiency is a quite important but hard to measure parameter. In theory the calculation is simple. One takes the number of multiplied electrons at the anode and divides it by the number of photoelectrons one originally had at the photo cathode (PC). The first can be calculated by using Poissonian statistics on a charge distribution and/or by using an adequate fit. The original number of photoelectrons at the PC can only be measured indirectly. The sensor is  illuminated in the QE-setup (200$\,$V between PC and the input of the MCPs) with a laser at a high frequency (10-50 MHz) but low number of photons per pulse ($\sim$ 5) and the corresponding photo current is measured. In a second measurement the laser frequency is reduced ($\sim$10 kHz) and the tube setup changed to the normal operating setup to measure the number of photoelectrons. In both setups a beam splitter is used to additionally measure the laser intensity with a calibrated photodiode since the number of photons per laser pulse is not stable for different frequencies. With this correction one gets:

\begin{equation}
CE = \frac{N_{pe,\,Anode,\,f_{low}}\cdot e \cdot f_{high}}{I_{PC,\,f_{high}}} \cdot \frac{I_{Diode,F_{high}}\cdot f_{low}}{I_{Diode,\,f_{low}}\cdot f_{high}} 
\end{equation}

In table \ref{tbl:CE} the measured CE values for different MCP-PMTs are listed. For a standard Photonis Planacon $\sim$65\% seems reasonable. Photonis 9002108 was the first Hi-CE tube we received and measured. Then many others followed and reached similar CEs like 9002192 and 9002193 ($\sim$ 88\%) or 9002220. Photek tubes can reach similarly high CEs as well. Hamamatsu originally used a protection film in front of the MCPs to increase the lifetime but this significantly reduced the CE so it was removed in the most recent tube again.

   \begin{table*}
     \centering
     \begin{tabular}{|c|c|c|c|c|c|c|c|} \hline 
       \makecell{Photonis \\ XP85112 \\ 9001394} & \makecell{Photonis \\ XP85112 \\ 9002108} & \makecell{Photonis \\ XP85112 \\ 9002192\&\\9002193} & \makecell{Photonis \\ XP85112 \\ 9002220} & \makecell{Photek \\ MAPMT253 \\ A1200116} & \makecell{Hamamatsu \\ R13266-07-64 \\ JS0022} & \makecell{Hamamatsu \\ R13266-07-64M \\ YH0250} \\ \hline 
     118\,M$\Omega$  & 10\,M$\Omega$ &  37\,M$\Omega$ \& 45\,M$\Omega$ & 37\,M$\Omega$ &  15\,M$\Omega$ & not measured & 53\,M$\Omega$  \\ \hline
             non-ALD      & ALD, first Hi-CE & ALD, Hi-CE & ALD, Hi-CE & ALD & \makecell{ALD, film in \\ front of MCPs}  & ALD, no film            \\ \hline 
        $(63\pm 6) \%$    & $(95\pm 9) \%$ & $(88\pm 8) \%$ & $(95\pm 5) \%$ & $(90\pm 9) \%$ & $(39\pm 4) \%$ & $(65\pm 7) \%$             \\ \hline 
      
     \end{tabular}

     \caption{resistance and collection efficiency for different MCP-PMTs}
     \label{tbl:CE}

   \end{table*}
   
\begin{figure}[h]
	\centering
				\subfloat{
			\includegraphics[trim = 0mm 0mm 0mm 15mm, clip =true, width=0.25\textwidth ]{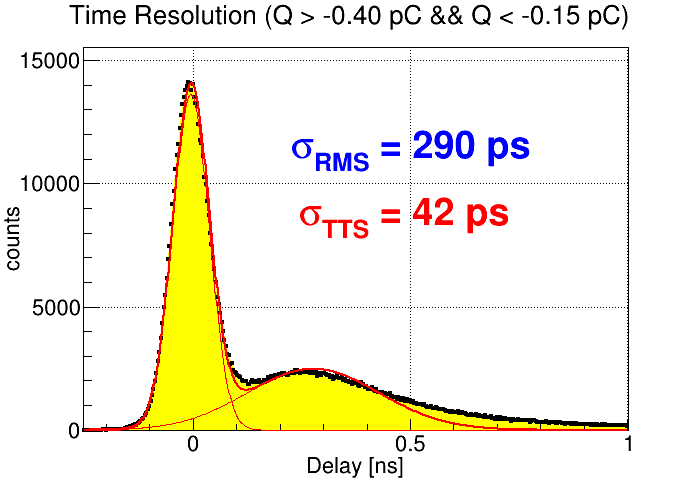}
			\includegraphics[trim = 0mm 0mm 0mm 15mm, clip =true, width=0.25\textwidth ]{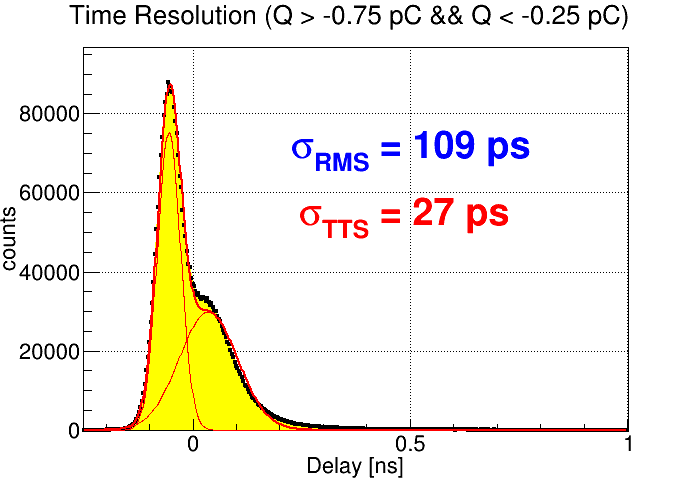}
		}	
	\caption{time resolution of Photonis 9002193 with 1:10:1 (left) and 4:10:1 (right) voltage divider}
	\label{fig:timeres}
\end{figure}

Unfortunately, this increase of the CE for Photonis MCP-PMTs had a negative effect on the time resolution. The high collection efficiency technique leads to more captured recoil electrons which come later in time. This leads to a worse $\sigma_{RMS}$ in the time window of $-0.5...2\,$ns. Most of the original time resolution can be regained by increasing the potential between the PC and the MCPs which can be done by adjusting the voltage dividers. The maxmimum is 4:10:1 (PC-MCPin:MCPin-MCPout:MCPout-Anode) from 1:10:1 which leads to a $\sigma_{RMS}$ of 109$\,$ps instead of 290$\,$ps as shown for an example sensor in figure \ref{fig:timeres} . 

\subsection{Gain homogeneity and gain inside B-fields}
\begin{figure}[h]
	\centering
	
				\subfloat{
			\includegraphics[trim = 0mm 0mm 0mm 10mm, clip =true, width=0.23\textwidth ]{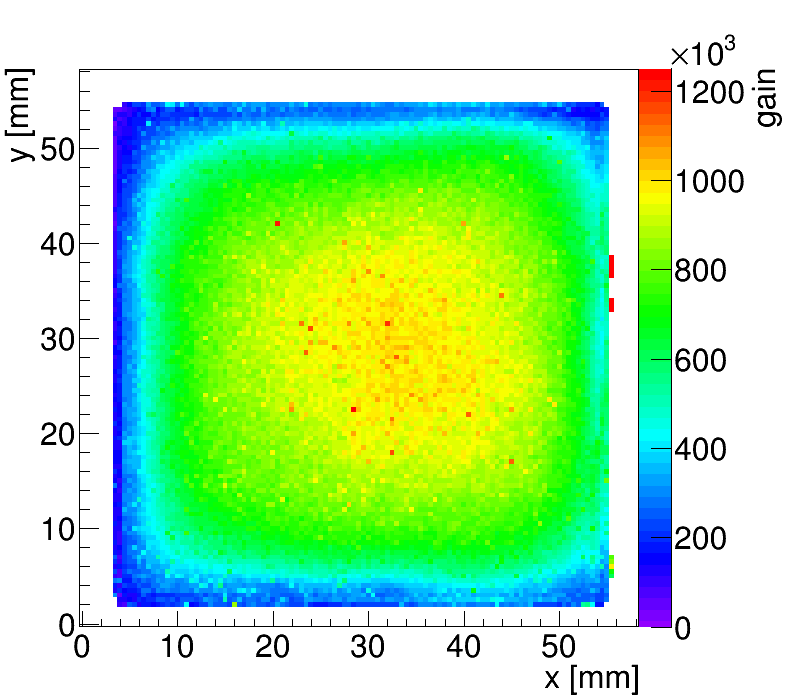}
			\includegraphics[trim = 0mm 0mm 0mm 10mm, clip =true, width=0.23\textwidth ]{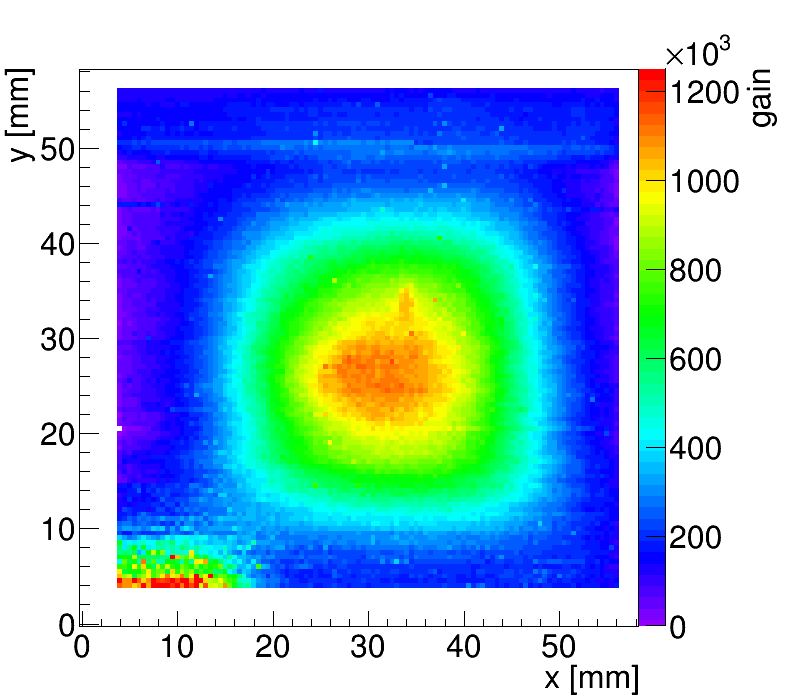}
		}
	\caption{gain scans of Photonis 9002192 (left) and Photek A3191220 (right)}
	\label{fig:gainscan}
\end{figure}

Another important parameter is the gain, both the absolute value (in magnetic fields as well) and its homogeneity across the sensor surface. The latter can be derived by scanning across the surface and measuring the current of all shorted anode pixels while illuminating one spot on the sensor. The result is a map of the relative gain folded with the QE so to get the true gain one has to divide by the map of the QE-scan. If one additionally measures the absolute gain at one point, for example by fitting a charge spectrum measured with a scope, the relative gain map can be scaled to the absolute gain map. In figure \ref{fig:gainscan} two gain scans are shown. Photonis 9002192 on the left side is quite homogeneous, Photek A3191220 on the right side is not. It also has an artificially high gain in the bottom left corner. The reason is the division with the almost zero QE in this area.

As already mentioned above the gain behaviour in magnetic fields is crucial as well. If the magnetic field is very high the electron multiplication in the MCPs ceases because the electrons start to curl around the magnetic field lines and won't hit the MCP pores' walls anymore.

\begin{figure}[h]
	\centering	
			\includegraphics[width=0.4\textwidth ]{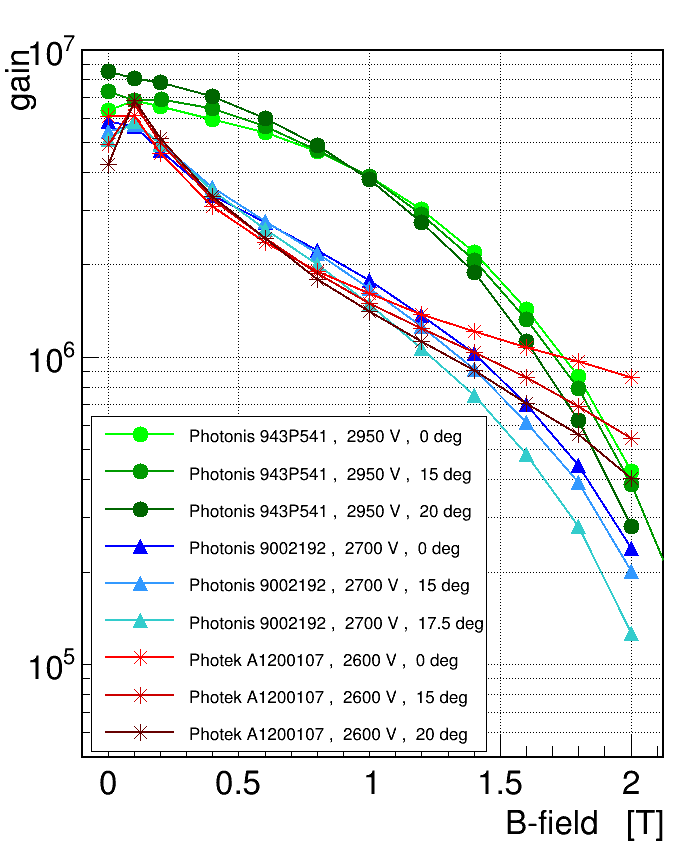}
	\caption{gain of three different MCP-PMTs Photonis 943P541 (3x100 pixels, 10$\,$\textmu m pores green), Photonis 9002192 (8x8 pixels, 10$\,$\textmu m pores, blue) and Photek A1200107 (8x8 pixels, 6$\,$\textmu m pores, red)}
	\label{fig:bfeld}
\end{figure}

Figure \ref{fig:bfeld} shows the gain vs. B-field relation for different MCP-PMTs under different tilting angles between PMT-axis and B-field direction. One can see a factor of $\sim$3 gain loss from zero to one Tesla, a little bit less for Photonis 946P541. At higher magnetic field the Photek tube shows a superior behaviour due to its 6$\,$\textmu m pores. Nevertheless all tubes reach 10$^6$ gain at 1$\,$T and 15 degrees tilt angle which is the requirement for the PANDA DIRCs.

\subsection{Darkcount rate, afterpulsing and charge sharing}
\begin{figure}[h]
	\centering
	\parbox{0.49\textwidth}{
	\centering
	\subfloat[]{
			\includegraphics[trim = 0mm 0mm 0mm 0mm, clip =true, width=0.49\textwidth ]{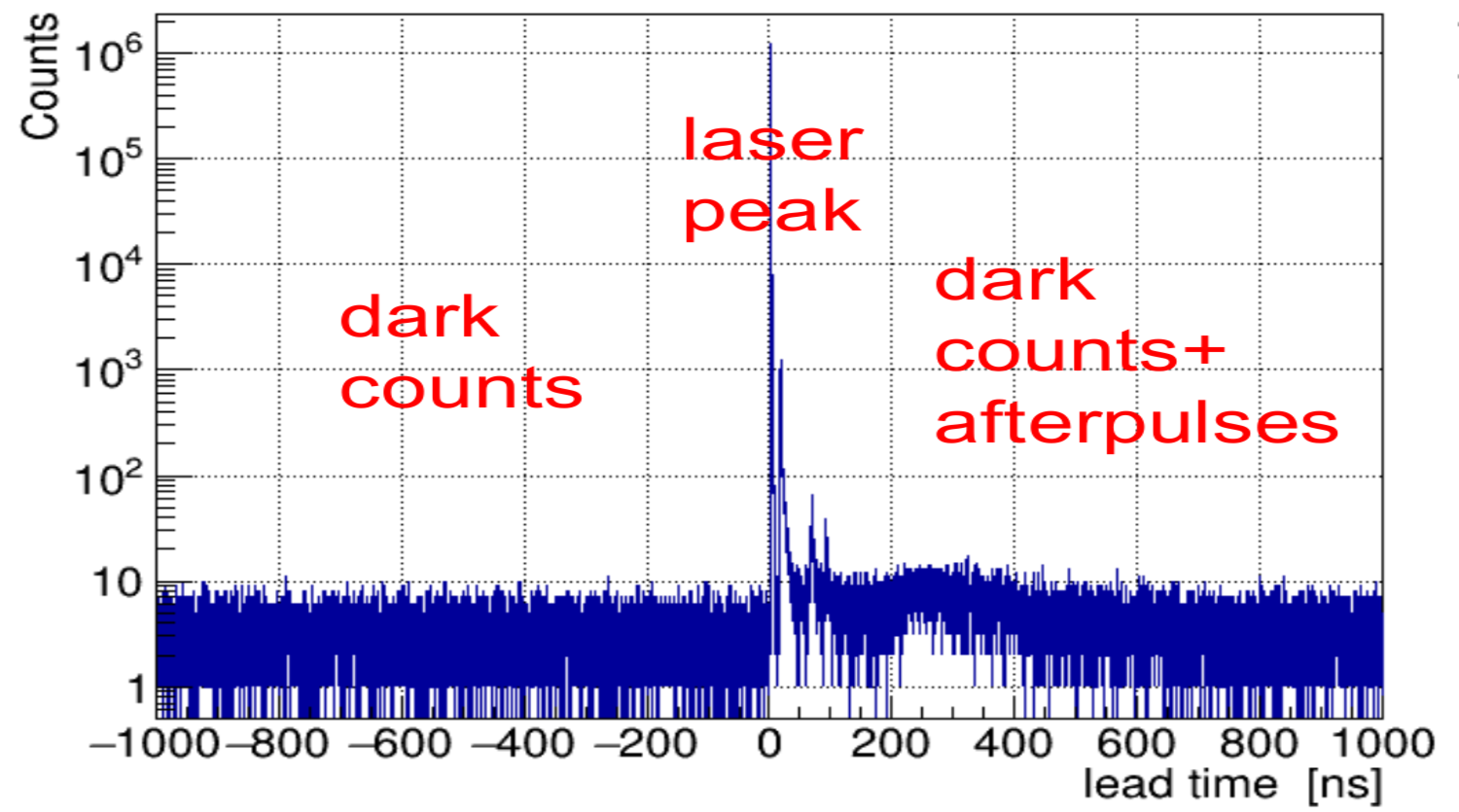}
			\label{fig:trbtime}
			}}
			\subfloat[]{
			\includegraphics[trim = 0mm 0mm 1mm 15mm, clip =true, width=0.23\textwidth ]{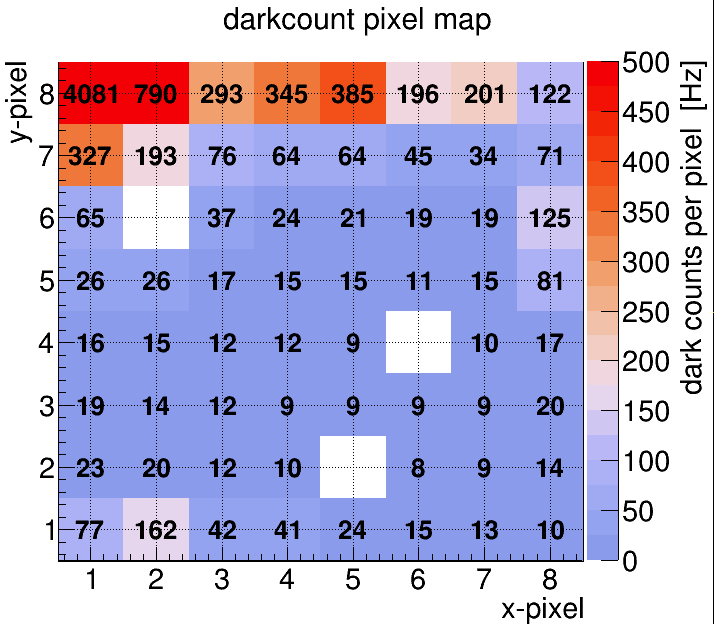}
			\label{fig:dc}}
			\subfloat[]{
			\includegraphics[trim = 1mm 0mm 0mm 15mm, clip =true, width=0.23\textwidth ]{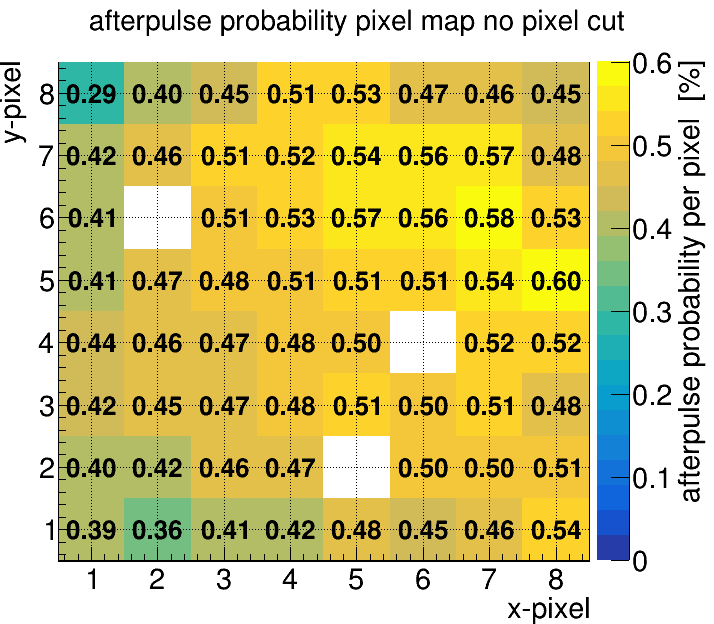}
			\label{fig:ap}
		}	
	\caption{example time spectrum, dark count map and afterpulse probability map of a Photonis 9002193 TRB-Scan}

\end{figure}

Scanning the sensor with a combination of PaDiWa or DiRICH frontend boards and TRB boards (in principle discriminator + TDC boards, for more information see: \cite{Ugur_Traxler_12_TRB} and \cite{Albert_18_TRB}) as DAQ gives some first level information like: x-y position, hit time (see figure \ref{fig:trbtime} for an example of a time spectrum for one pixel of Photonis 9002193), time-over-threshold (ToT) and the number of hits per laser trigger. A combination of them yields (among others) higher level information \cite{EDIT_2010} like the time resolution, the dark count rate distribution (figure \ref{fig:dc}), charge sharing (and electronic) crosstalk distribution, recoil electron distributions (both in space and time) and afterpulse probability distributions (figure \ref{fig:ap}). The dark count distribution looks quite typical, mostly homogeneous and low with a few hot pixels on the edges or in the corners up to 4\,kHz in the top left corner. The afterpulse probability is quite low with about half a percent across the sensor surface.

\subsection{Rate capability}

\begin{figure}[h]
	\centering	
			\includegraphics[width=0.49\textwidth]{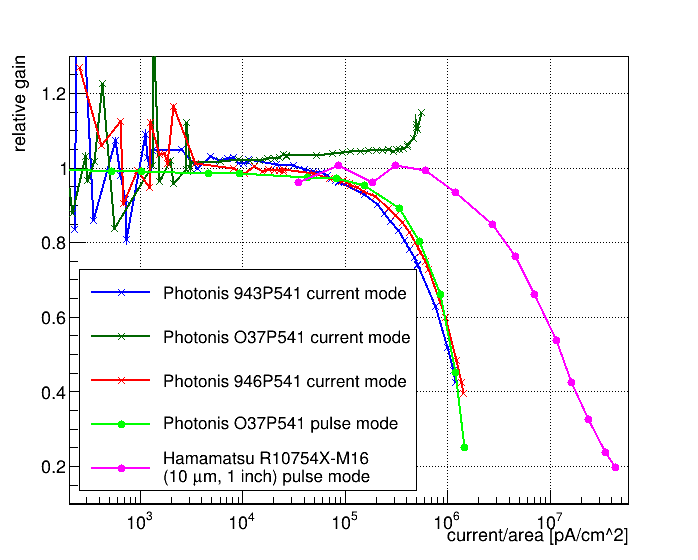}
	\caption{rate capability of different MCP-PMTs measured with different techniques}
	\label{fig:ratestab}
\end{figure}

Due to the high PANDA interaction rate the tubes must withstand high photon rates as well. One possibility to measure the rate capability is to illuminate a pixel and measure its gain with increasing laser frequcency (=pulse mode). Another way is to short all anode pixels and measure the resulting current (=current mode). If it increases slower than the light intensity (for example measured with a photo diode), then the MCP-PMT saturates.
Figure \ref{fig:ratestab} shows the rate capability of some 53x53$\,$mm$^2$ tubes and a Hamamatsu 26x26$\,$mm$^2$ for comparison. For the PANDA EDD about $1\,$MHz/cm$^2$ of photo electrons are expected which corresponds to $\sim$2$\cdot 10^5\,$pA/cm$^2$ at a gain of 10$^6$. For the Barrel DIRC 5x fewer photo electrons are anticipated. Therefore all 53x53$\,$mm$^2$ meet the requirements of both DIRCs.
If one looks at the data of the Photonis O37P541 in current mode, an unexpected behavior is observed: the gain shows an increase instead of a decline. This is part of a strange effect which we call "escalation" which is discussed next.

\subsection{"Escalation"}

\begin{figure}[h]

	\centering
	\subfloat{
				\includegraphics[trim = 0mm 0mm 0mm 15mm, clip =true, width=0.24\textwidth ]{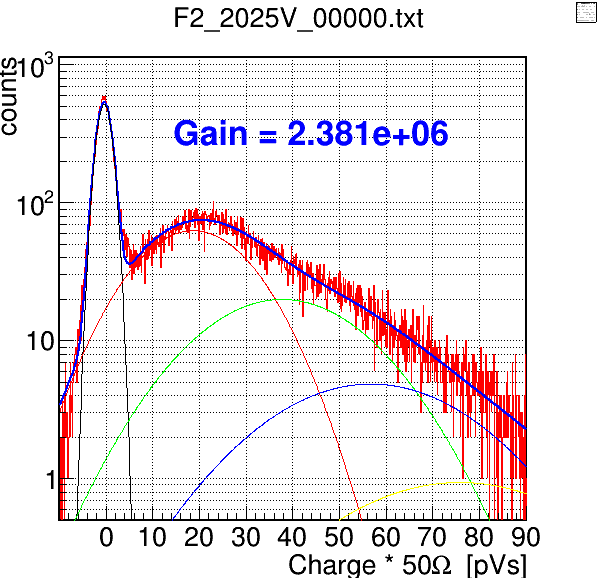}

			\includegraphics[trim = 0mm 0mm 0mm 15mm, clip =true, width=0.24\textwidth ]{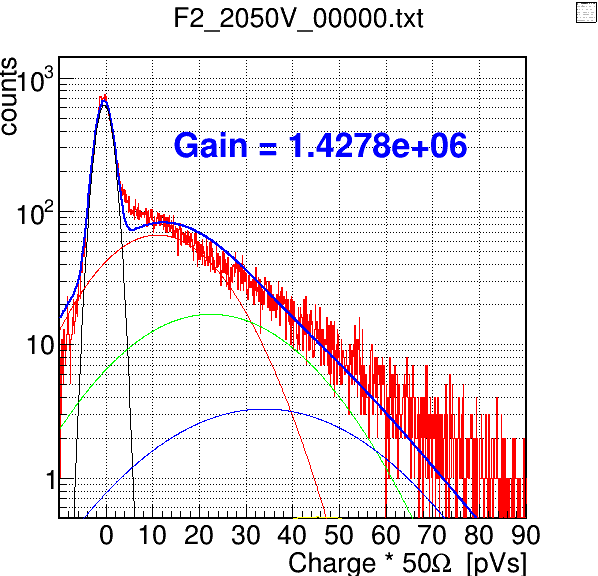}
			
		}
			
	\caption{charge spectrum of Photonis O37P541 at 2025$\,$V (just before escalation, left) and at 2050$\,$V in escalation mode (right)}
	\label{fig:esca}

\end{figure}

\begin{figure}[h]

	\centering
	\subfloat{
	
					\includegraphics[trim = 20mm 20mm 40mm 15mm, clip =true, width=0.24\textwidth ]{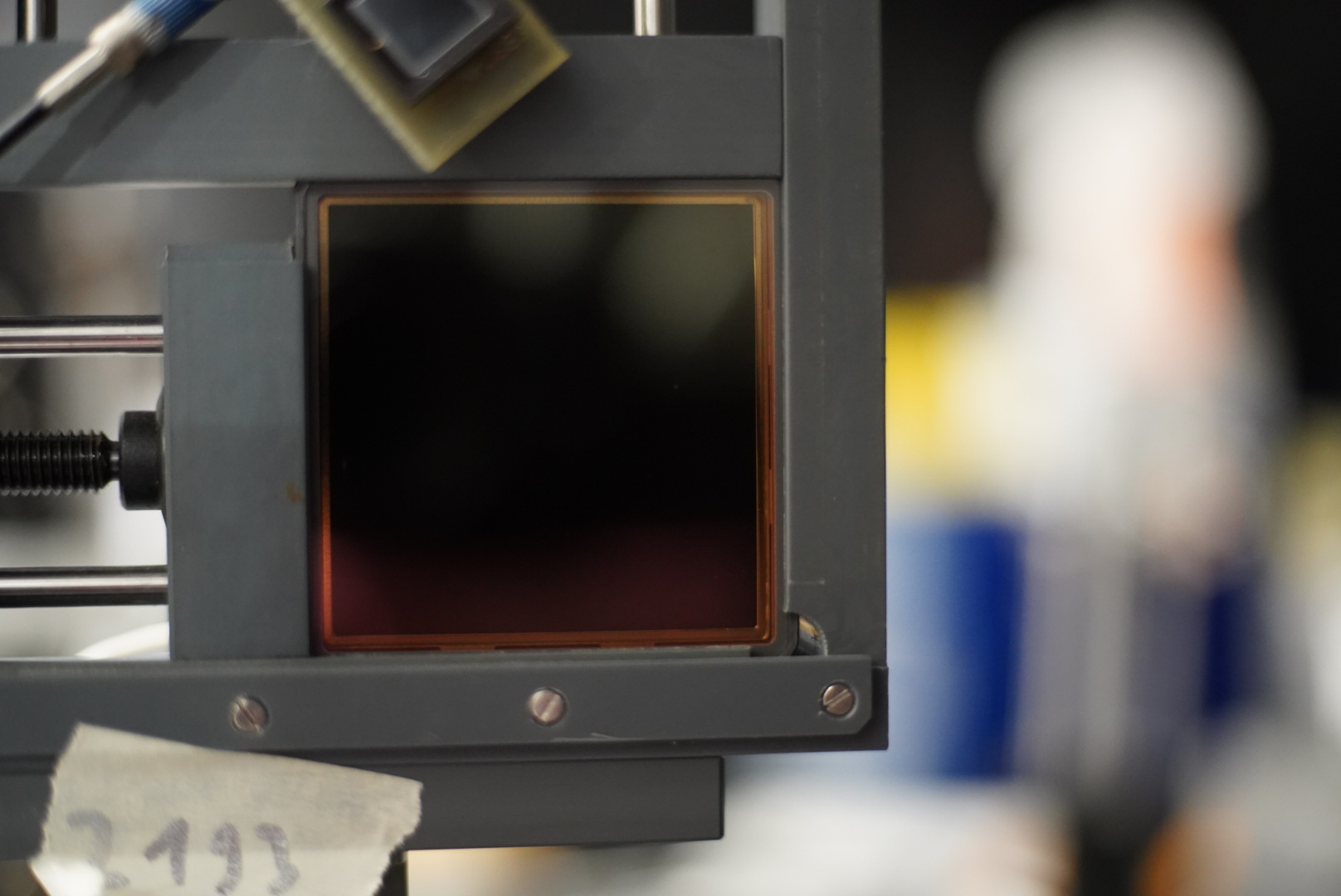}

			\includegraphics[trim = 20mm 20mm 40mm 15mm, clip =true, width=0.24\textwidth ]{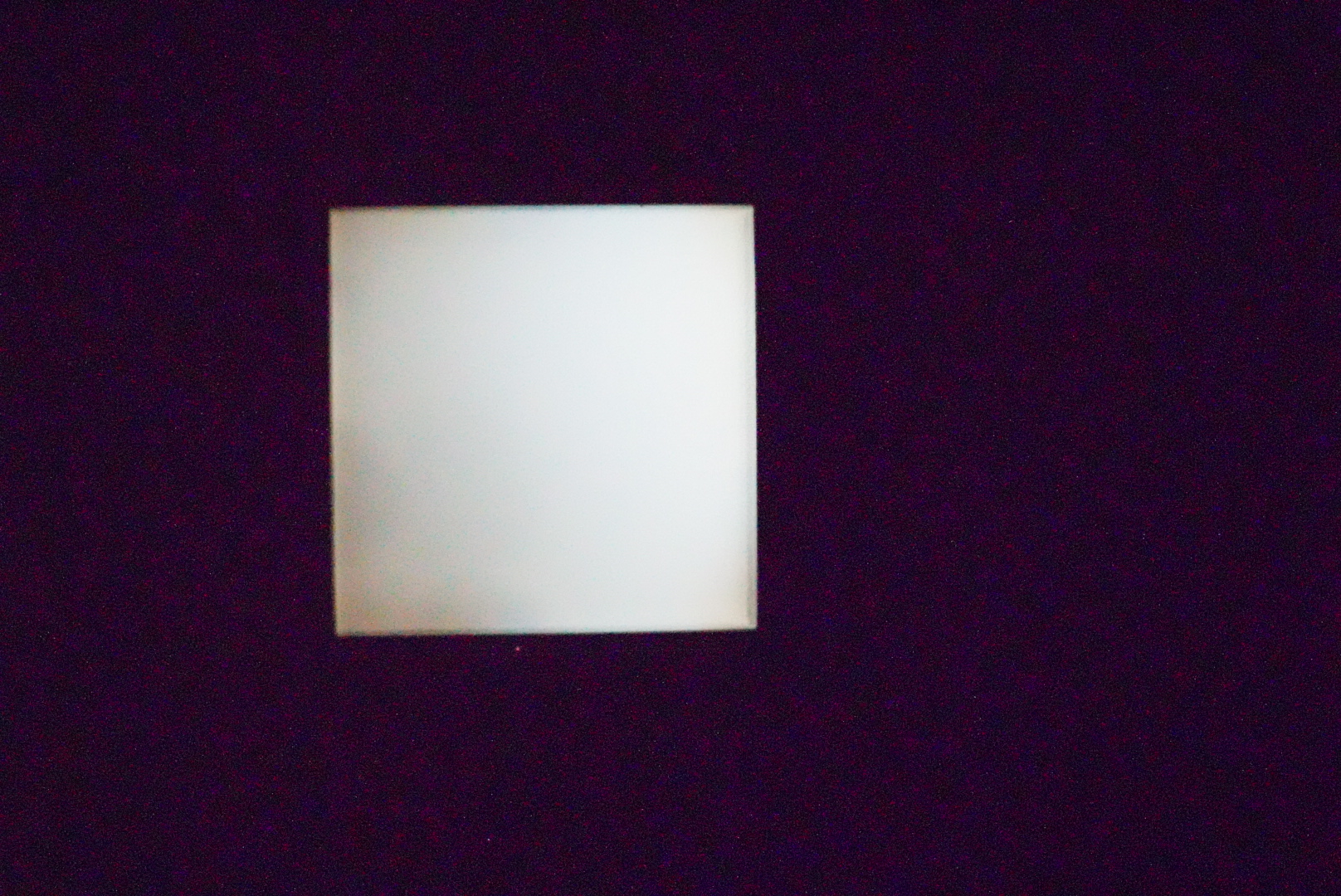}
			
			}
			
	\caption{photo of Photonis 9002193 with light on (left) and with light off in escalation mode (right)}
	\label{fig:licht}

\end{figure}

There is an odd state which some MCP-PMTs can enter. At the moment only the latest Photonis tubes show this effect. This can happen when operating the tube at high voltages (and thus gain) or very high illumination (as in the rate capability measurement). Several things happen simultaneously when the tube goes into this state: One can see a very high (dark) count rate and the overall signal size and gain drop. Also the charge spectrum looks distorted (see figure \ref{fig:esca}) The anode current increases drastically and also the current across the MCPs increases suddenly by a factor of $\sim$3 and then steadily increases further, while the MCP resitivity drops.
Inside a magnetic field this effect seems to be less serious and appears later. 
The origin of this effect seems to be photon creation inside the MCP-PMT and tests showed that the emitted photons originate in the MCPs.
Under certain conditions there are so many photons emitted that one can take a photo or even see the light by bare eye (see figure \ref{fig:licht}). These photons bring the MCP-PMT to saturation (high count rate, but low gain and high anode current).\\
It seems likely that during the electron amplication process there are photons emitted from the ALD layers. These photons release further photo electrons from the PC and at high gain and/or at high illumination. This causes a  kind of self-stimulated photon emission and the system runs away.
 It is not yet understood why the current across the MCPs increases, where the sensitivity to magnetic fields comes from and how the mechanism works in detail. Photonis is already aware of this effect and working on MCPs that enter this state later or in best case not at all.

\section{Conclusions and outlook}

There were significant improvements in the performance of MCP-PMTs in the recent years, especially regarding the lifetime with the ALD-coating and in collection efficiency. Most of the recent tubes are suited for operation in both PANDA DIRCS. For the Barrel DIRC 155 53x53 mm$^2$ ALD-coated MCP-PMTs with 10\,\textmu m pores were ordered from Photonis. They will start with the mass prodcution when the "escalation" problem is solved or at least sufficiently suppressed. The tubes will be delivered to Erlangen for the quality assurance check measurements and then be built installed into PANDA.

\section*{Acknowledgments}

We thank our Erlangen colleagues of ECAP for granting us use of the QE setup. 
This work is supported by the German BMBF and GSI Darmstadt.


\bibliography{mybibfile}

\end{document}